\documentclass[preprint,english,prl,amsmath,showpacs,amssymb]{revtex4}
\usepackage{graphicx,esint,bm,float,lipsum,babel}
\makeatletter
\@ifundefined{textcolor}{}
{%
 \definecolor{BLACK}{gray}{0}
 \definecolor{WHITE}{gray}{1}
 \definecolor{RED}{rgb}{1,0,0}
 \definecolor{GREEN}{rgb}{0,1,0}
 \definecolor{BLUE}{rgb}{0,0,1}
 \definecolor{CYAN}{cmyk}{1,0,0,0}
 \definecolor{MAGENTA}{cmyk}{0,1,0,0}
 \definecolor{YELLOW}{cmyk}{0,0,1,0}
 }
\makeatother

\begin{document}

\title{Origin  of electronic Raman scattering and the Fano resonance\\
 in metallic carbon nanotubes}
\author{Eddwi~H.~Hasdeo$^{1}$}\email [E-mail:
]{hasdeo@flex.phys.tohoku.ac.jp}
\author{Ahmad~R.~T.~Nugraha$^{1}$}
\author{Riichiro~Saito$^{1}$}
\author{Kentaro~Sato$^{1}$}
\author{Mildred~S.~Dresselhaus$^{2,3}$}

\affiliation{$^{1}$Department of Physics, Tohoku University, Sendai
  980-8578, Japan\\
  $^{2}$Department of Physics, Massachusetts Institute of Technology,
  Cambridge, MA 02139-4307, USA\\
  $^{3}$Department of Electrical Engineering, Massachusetts Institute
  of Technology, Cambridge, MA 02139-4307, USA}

\date{\today}

\begin{abstract}
  Fano resonance spectra for the G band in metallic carbon nanotubes
  are calculated as a function of laser excitation energy in which the
  origin of the resonance is given by an interference between the
  continuous electronic Raman spectra and the discrete phonon spectra.
  We found that the second-order scattering process of the ${\bf
    q}\neq0$ electron-electron interaction is more relevant to the
  continuous spectra rather than the ${\bf q}=0$ first-order process
  because the ${\bf q}=0$ direct Coulomb interaction vanishes due to
  the symmetry of the two sublattices of a nanotube.  We also show
  that the RBM spectra of metallic carbon nanotubes have an asymmetric
  line shape which previously had been overlooked.
\end{abstract}


\maketitle
Raman spectroscopy of single wall carbon nanotubes (SWNTs) and
graphene has provided us with a better understanding of many optical
properties which are very important for characterizing SWNTs and
graphene not only for basic science understanding but also in
applications~\cite{doi:10.1080/00018732.2011.582251}.  Although most
of the excitonic physics in the Raman spectra of SWNTs has been
investigated intensively in terms of, for example, the excitation
energy dependence (resonance
Raman)~\cite{jorio01-raman,Pimenta_resonantpapertua}, chirality
dependence (the Kataura
plot)~\cite{kataura99-plot,saito00-trig,jiang07-exc}, Fermi energy
dependence (the Kohn
anomaly)~\cite{PhysRevLett.99.145506,sasaki08-rbm,PhysRevB.80.081402},
polarization
dependence~\cite{PhysRevB.74.155411,PhysRevB.74.205440,PhysRevB.67.165402},
and even strain
dependence~\cite{Li20114412,Cronin_RRS_axial_strain},
however, the asymmetric spectral shape of the G band for metallic
SWNTs (m-SWNTs), also known as Breit-Wigner-Fano (BWF) line shape, is
still not well explained theoretically.  In a previous study, Brown
\emph{et al}. showed the diameter-dependent asymmetric spectral shape
of the G band in which the asymmetric factor $1/q_{\mathrm{BWF}}$
depends on the density of states at the Fermi
energy~\cite{PhysRevB.63.155414}. Additionally, the BWF line shapes
appear in graphite intercalation compounds (GICs) where
$1/q_{\mathrm{BWF}}$ depends on the staging number of GICs and thus
also depends on the density of states at the Fermi
energy~\cite{PhysRevB.16.3330}.  Therefore, electrons in the gapless
linear energy band of m-SWNTs should be expected to exhibit these
asymmetry-related phenomena.

Fano pointed out that the asymmetric feature of a broadened spectrum
comes from an interference between a discrete excitation spectrum and
a continuum spectrum~\cite{PhysRev.124.1866}.  In m-SWNTs, electrons
in the linear energy band play an important role to give rise to the
continuum spectra and phonons give the discrete spectra.  However, the
detailed mechanism of the BWF line shapes in m-SWNTs remains a
long-standing debatable topic. Some reports suggest that the coupling
of a collective excitation (plasmon) with a phonon could explain the
origin of the BWF asymmetry~\cite{PhysRevB.72.153402,
  PhysRevB.66.161404, PhysRevB.66.195406, PhysRevB.63.155414}, and
some others argue that the single-particle electron-hole pair and
phonon coupling via the Kohn anomaly is more
relevant~\cite{Lazzeri-no-plasmon,Wu_variable-eph}.  Recently,
Farhat~\emph{et al.} have observed a new feature of the continuum
spectra exclusively in m-SWNTs which is ascribed to the electronic
Raman scattering (ERS)~\cite{PhysRevLett.107.157401}.  The ERS feature
(at $\sim$500 $\mathrm{cm^{-1}}$) is observed in the energy region
between the RBM and the G band and shows no phonon feature by the
following arguments: (1) in comparison to the phonon spectral width
($\sim$1-50 $\mathrm{cm^{-1}}$), the ERS width is much broader
($\sim$500 $\mathrm{cm}^{-1}$) and has a smaller peak intensity
($I_{\mathrm{ERS}}\approx0.6I_{\mathrm{G}}$), (2) the energy of the
inelastic scattered light in the phonon Raman spectra is changed by
changing the laser excitation energy $E_{\rm{L}}$, while the ERS peak
position does not change; it keeps constant at $M_{ii}$ ($i^{th}$ Van
Hove singularity transition energy), (3) the ERS feature is suppressed
by changing the Fermi energy, which indicates that the origin of this
spectrum comes from electron-hole pair excitations in the linear band
of m-SWNTs by the Coulomb interaction.

In this Letter, we propose that the BWF feature of m-SWNTs comes from
the interference between the G band and the ERS spectra.  We calculate
the exciton-exciton matrix elements of the Coulomb interaction which
are responsible to give the ERS spectra. The exciton consideration is
based on the fact that exciton effects in m-SWNTs are not negligible
due to the one-dimensional carrier confinement even in the presence of
the screening effect~\cite{PhysRevLett.99.227401,
  PhysRevLett.92.077402}.  The calculated results of the present work
suggest that the zero momentum transfer (${\bf q}=0$) vanishes in the
direct Coulomb interaction because of the symmetry of the
wave-function, and thus a higher order Raman process is more relevant
to the ERS. By considering the second-order Raman process, we are able
to reproduce experimental results of the ERS spectra consistently.  We
will also show that the RBM spectra of m-SWNTs have a similar
asymmetric line shape, indicating that the ERS can be coupled with
both the G band and the RBM.

\begin{figure}
\includegraphics[clip,width=\columnwidth]{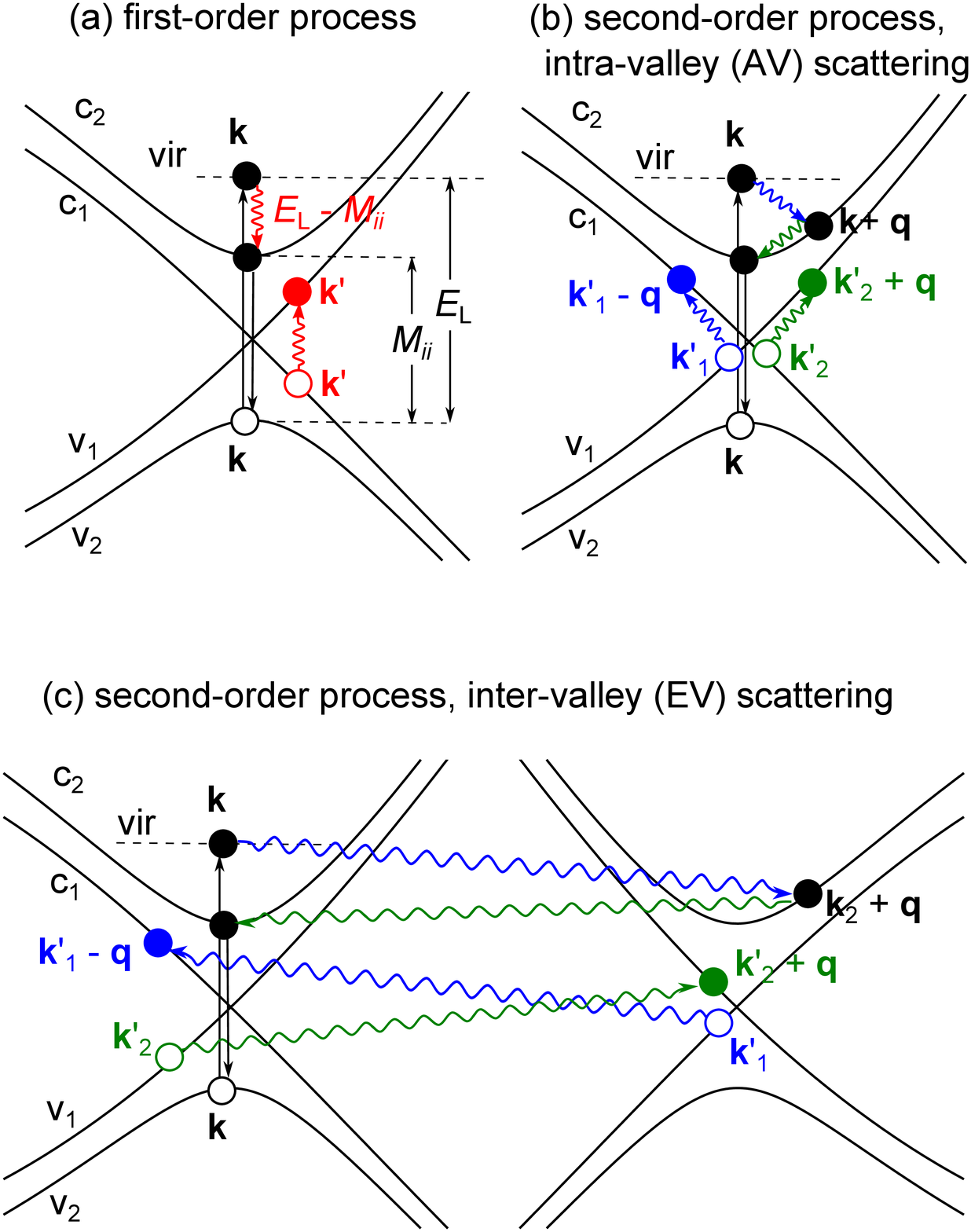}
\caption{\label{fig:ersprocess}(Color online) (a) First-order
  electronic Raman process.  (b) AV and (c) EV second-order scattering
  processes $\left({\bf q}\ne0\right)$.  In both the first-order
  and second-order processess, the interaction between electrons in
  the parabolic band and the linear band can take place in the same
  valley ($\mathbf{K}$ or $\mathbf{K'}$) or in a different valley. }
\end{figure}

Optical processes of the ERS consist of (i) an exciton generation via
an exciton-photon interaction, (ii) excitation of another exciton in
the linear energy band by the Coulomb interaction with the
photo-excited exciton, and (iii) finally the photo-excited exciton
goes back to the ground state by emitting a photon.  The
exciton-exciton interaction in (ii) may occur in a first-order or
high-order process. Here, we consider up to second-order processes for
simplicity.  For the first-order process, the photo-excited exciton
relaxes vertically (${\bf q}=0$) from a virtual state $\Psi^{\rm{vir}}$  to
the $M_{ii}$ state after photo-absorption at a wave vector ${\bf k}$,
while the other exciton is created in the linear band at wave vector
${\bf k}^{'}$ by the Coulomb interaction (see
Fig.~\ref{fig:ersprocess}(a)). In the second-order process, on the
other hand, the existence of the two inequivalent $\mathbf{K}$ and
$\mathbf{K}^{'}$ points in the graphene Brillouin zone leads to two
different scattering processes, i.e.~\emph{intra-valley} (AV)
scattering and the \emph{inter-valley} (EV) scattering, shown in
Figs.~\ref{fig:ersprocess}(b) and (c), respectively. In both cases,
two excitons are excited at the linear band.  For each scattering
process, we also have two cases where the two electrons at parabolic
and linear bands may exist in the same valley ($A$ state), or they may
exist in the different valleys ($E$ state).  Such a symmetry labeling for $A$
and $E$ states is obtained from the group theory.  After going through
the electronic scattering process, the photo-excited exciton then
returns to the ground state by emitting a photon with resonance energy
$E_{\mathrm{s}}=M_{ii}$.  This is the reason why the ERS peak remains
at $M_{ii}$ even though we change the laser energy $E_{{\rm L}}$.

Considering all the processes shown in Fig.~\ref{fig:ersprocess}, we
write the perturbed Hamiltonian as:
\begin{align}
  H_{\mathrm{e-e}} =& \sum_{\mathbf{k},\mathbf{k}^{'},\mathbf{q}}
  W^{(\pm)}\left(\left(\mathbf{k}+\mathbf{q}\right),
    \left(\mathbf{k}^{'} - \mathbf{q}\right){},
    \mathbf{k},\mathbf{k}^{'} \right)\nonumber\\
  & \times c_{\mathbf{k}+\mathbf{q}}^{\dagger\mathrm{c}}
  c_{\mathbf{k}^{'}-\mathbf{q}}^{\dagger\mathrm{c'}}c_{\mathbf{k}^{'}}^{\mathrm{v'}}
  c_{\mathbf{k}}^{\mathrm{c}},\label{eq:hee}
\end{align}
where $\mathbf{k}$ and $\mathbf{k}^{'}$ denote, respectively, an
electron state in the parabolic and the linear band, while
$c_{\mathbf{k}}^{\dagger\mathrm{c}}$$\left(c_{\mathbf{k}}^{\mathrm{v}}\right)$
is the creation (annihilation) operator in the conduction (valence) band.
The direct (exchange) interaction $K^{\mathrm{d}}$ ($K^{\mathrm{x}}$)
contributes to the two-body Coulomb interaction $W$ as follows:
$W{}^{(\pm)}=K^{\mathrm{d}}\pm K^{\mathrm{x}}$, in which
$+\left(-\right)$ gives a singlet (triplet) state for the two
electrons. $K^{\mathrm{d}}$ and $K^{\mathrm{x}}$ are
expressed as \cite{supplement}:
\begin{eqnarray}
  K^{\mathrm{d}} & = &
  \sum_{ss'=\rm{A},\rm{B}} C_{s,\mathbf{k}+\mathbf{q}}^{\mathrm{c}*}
  C_{s',\mathbf{k^{'}}-\mathbf{q}}^{c*}
  C_{s,\mathbf{k}}^{\mathrm{c}} C_{s',\mathbf{k}^{'}}^{\mathrm{v}}\nonumber\\
  & &
  \times\Re\left(w_{ss'}\left(\mathbf{q}\right)\right),\label{eq:Kd}\\
  K^{\mathrm{x}} & = &
  \sum_{ss'=\rm{A},\rm{B}}C_{s,\mathbf{k}+\mathbf{q}}^{c*}C_{s',\mathbf{k^{'}}-\mathbf{q}}^{c*}
  C_{s',\mathbf{k}}^{\mathrm{c}}C_{s,\mathbf{k}^{'}}^{\mathrm{v}}\nonumber
  \\
  & &
  \times \Re \left(w_{ss'}\left(\mathbf{k}^{'} - \mathbf{k}-\mathbf{q}
    \right)\right), \label{eq:Kx}
\end{eqnarray}
where $C_{s,{\bf k}}^{{\rm c}({\rm v})}$ are the tight binding coefficients for
$s=$ A, B atomic sites of the conduction (valence) band, $\Re ()$ is the
real part of a complex variable, and the screened potential
$w\left(\mathbf{q}\right)$ is given by the random phase approximation
(RPA): $w(\mathbf{q}) = v(\mathbf{q})/\kappa
\left(1+v\left(\mathbf{q}\right)\Pi\left(\mathbf{q}\right)\right)$
\cite{jiang07-exc,PhysRevB.67.165402}.  Here
$v\left(\mathbf{q}\right)$ denotes the Fourier transform of the Ohno
potential, $\Pi\left(\mathbf{q}\right)$ is the RPA polarization
function, and $\kappa$ is the static dielectric constant due to
electronic core states, $\sigma$ bands, and the surrounding material.  In
this calculation we used a constant $\kappa=2.2$ \cite{jiang07-exc}.

The exciton-exciton matrix element for the photo-excited exciton and
another exciton in a linear energy band is calculated using the
following formula:
\begin{align}
  {\cal M}_{\rm{ex-ex}}^{\pm}\left(\mathbf{q}\right) =&
  \langle\Psi^{f}|H_{\mathrm{e-e}}|\Psi^{\mathrm{vir}}\rangle \nonumber\\
  =& \sum_{\mathbf{k},\mathbf{k^{'}}}
  Z_{\left(\mathbf{k}+\mathbf{q}\right)\mathrm{c},\mathbf{k}\mathrm{v}}^{*}
  Z^{*}{}_{\left(\mathbf{k}^{'}-\mathbf{q}\right)\mathrm{c},\mathbf{k}^{'}\mathrm{v}}
  Z_{\mathbf{k}\mathrm{c},\mathbf{k}\mathrm{v}} \nonumber\\
  & {} \times W^{(\pm)}\left(\left(\mathbf{k}+\mathbf{q}\right),
    \left(\mathbf{k}^{'}-\mathbf{q}\right){},
    \mathbf{k},\mathbf{k}^{'}\right).\label{eq:mxx}
\end{align}
Here the photo-excited  exciton state is defined by:
\begin{equation}
  |\Psi^{\mathrm{vir}}\rangle =
  \sum_{n, \mathbf{k}}Z_{\mathbf{k}\mathrm{c},\mathbf{k}\mathrm{v}}^{n}
  c_{\mathbf{k}}^{\dagger\mathrm{c}}
  c_{\mathbf{k}}^{\mathrm{v}}|g\rangle, \label{psiop}
\end{equation}
where $Z_{\mathbf{k}_{\mathrm{c}},\mathbf{k}_{\mathrm{v}}}^{n*}$ is
the eigenvector of $n\mathrm{-th}$ exciton state solved from the
Bethe-Salpeter equation, $\mathbf{k}_{c}$ and
$\mathbf{k}_{\mathrm{v}}$ denote wave vectors for the electron and hole
states, respectively, with $\mathbf{k}_{c}=\mathbf{k}_{\mathrm{v}}$
for a bright exciton, and $|g\rangle$ denotes the ground
state~\cite{jiang07-exc}.  In Eq.~\eqref{psiop}, we only use the
lowest exciton state $n = 0$, since it gives the dominant value to
the exciton-photon matrix element \cite{PhysRevB.75.035405}.The final
state of Eq. \eqref{eq:mxx} is given by:
\begin{equation}
  |\Psi^{f}\rangle =
  \sum_{\mathbf{k},\mathbf{k}^{'}}
  Z_{\left(\mathbf{k}+\mathbf{q}\right)\mathrm{c},\mathbf{k}\mathrm{v}}
  Z{}_{\left(\mathbf{k}^{'}-\mathbf{q}\right)\mathrm{c},\mathbf{k}^{'}\mathrm{v}} 
  c_{\mathbf{k}+\mathbf{q}}^{\dagger\mathrm{c}}
  c_{\mathbf{k}^{'}-\mathbf{q}}^{\dagger\mathrm{c}}
  c_{\mathbf{k}^{'}}^{\mathrm{v}} c_{\mathbf{k}}^{\mathrm{v}}|g\rangle.
\end{equation}

\begin{figure}
\includegraphics[clip,width=\columnwidth]{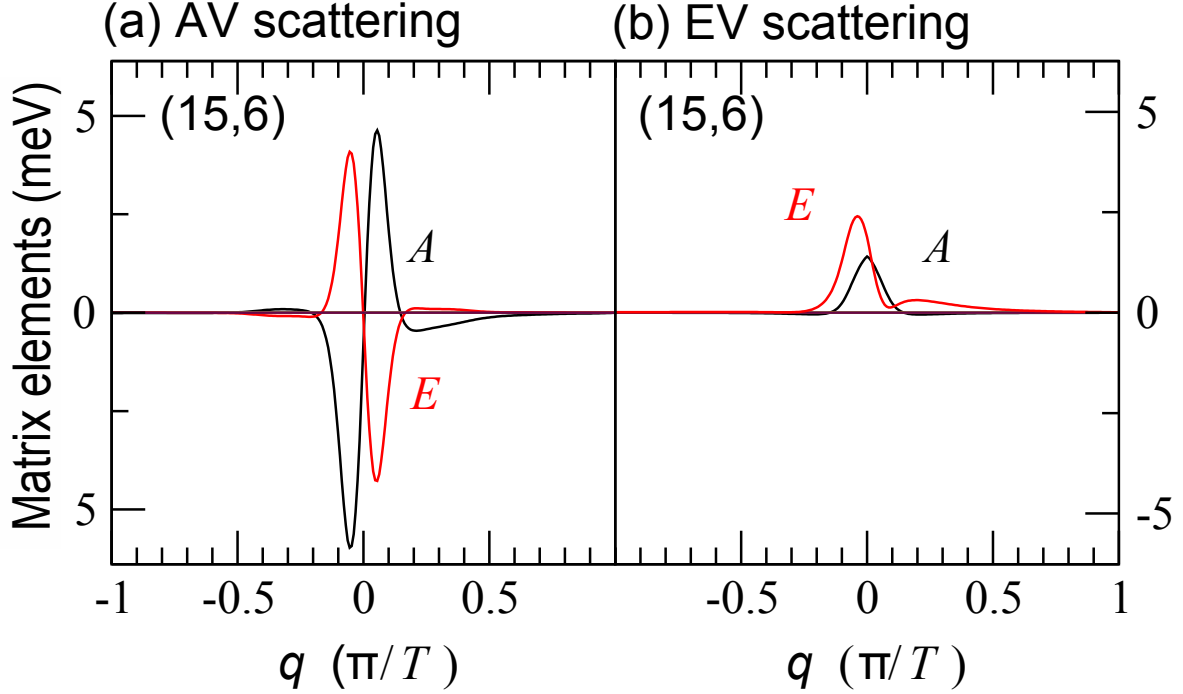}
\caption{\label{fig:Exciton-exciton-matrix-elements}(Color online)
  Singlet state exciton-exciton matrix elements $ {\cal
    M}_{\rm{ex-ex}}^{+}$ calculated for a (15,6) tube with a diameter
  of $1.46$ nm.  Panel (a) is for the intra-valley scattering and (b)
  is for the inter-valley scattering.  Label $A$ ($E$) inside each
  panel shows the $A$ ($E$) states, in which two electrons lie in the
  same (different) valley.  The wave vector ${\bf q}$ is projected on
  the one-dimensional SWNT cutting lines and expressed in terms of the
  translational vector length $T$.  We have $T = 0.89$ nm for the
  (15,6) tube.}
\end{figure}

In Fig.~\ref{fig:Exciton-exciton-matrix-elements}, we show the
calculated exciton-exciton matrix elements for singlet states $ {\cal
  M}_{\rm{ex-ex}}^{+}$ in a (15,6) m-SWNT.  It is noted that the
matrix elements for the triplet states $ {\cal M}_{\rm{ex-ex}}^{-}$
are comparable with those for the singlet states, so here we only show
the singlet state case.  Surprisingly, the AV scattering matrix
elements give almost a zero value at ${\bf q} = 0$ for both $A$ and
$E$ states, as shown in
Fig. \ref{fig:Exciton-exciton-matrix-elements}(a). In fact, we find
that the direct interaction $K^{\mathrm{d}}$ vanishes at
$\mathbf{q}=0$ for all nanotubes. At ${\bf q}=0$, only the exchange
interaction $K^{\mathrm{x}}$ gives a small contribution from the AV
scattering.  The vanishing $K^{\mathrm{d}}$ can be explained by the
presence of three $C^{\mathrm{c}}_s$ and one $C^{\mathrm{v}}_s$
coefficients in Eq.~\eqref{eq:Kd}.  The product of wave functions
always gives an opposite sign when we exchange A $\rightarrow$ B in
$s$ or $s'$ and thus the total summation over A and B sublattices
vanishes at ${\bf q}=0$ \cite{supplement}.  As long as we incorporate three
$C^{\mathrm{c}}$ and one $C^{\mathrm{v}}$ coefficients into
$K^{\rm{d}}$, the vanishing direct Coulomb interaction at ${\bf q} =
0$ is a general phenomenon in graphene and SWNTs systems.
Furthermore, the EV scattering matrix elements shown in
Fig.~\ref{fig:Exciton-exciton-matrix-elements}(b) are an even function
of $q$ because $C^{\mathrm{c}}$ and $C^{\mathrm{v}}$ change their
signs by exchanging $\mathbf{K}$ and $\mathbf{K'}$ in the B
sublattice, while in the A sublattice there are no changes in sign for
$C^{\mathrm{c}}$ and $C^{\mathrm{v}}$.  The results from
Figs.~\ref{fig:Exciton-exciton-matrix-elements}(a) and (b) thus imply
that the first-order Raman process corresponding to the AV scattering
at $\mathbf{q}=0$ as shown in
Fig.~\ref{fig:Exciton-exciton-matrix-elements}(a) makes only a minor
contribution to the Raman spectra.  Consequently, we should consider
the second-order ERS process, in which the ${\bf q} \neq 0$ term in
Fig.~\ref{fig:Exciton-exciton-matrix-elements}(a) becomes important.

\begin{figure}
\includegraphics[width=\columnwidth]{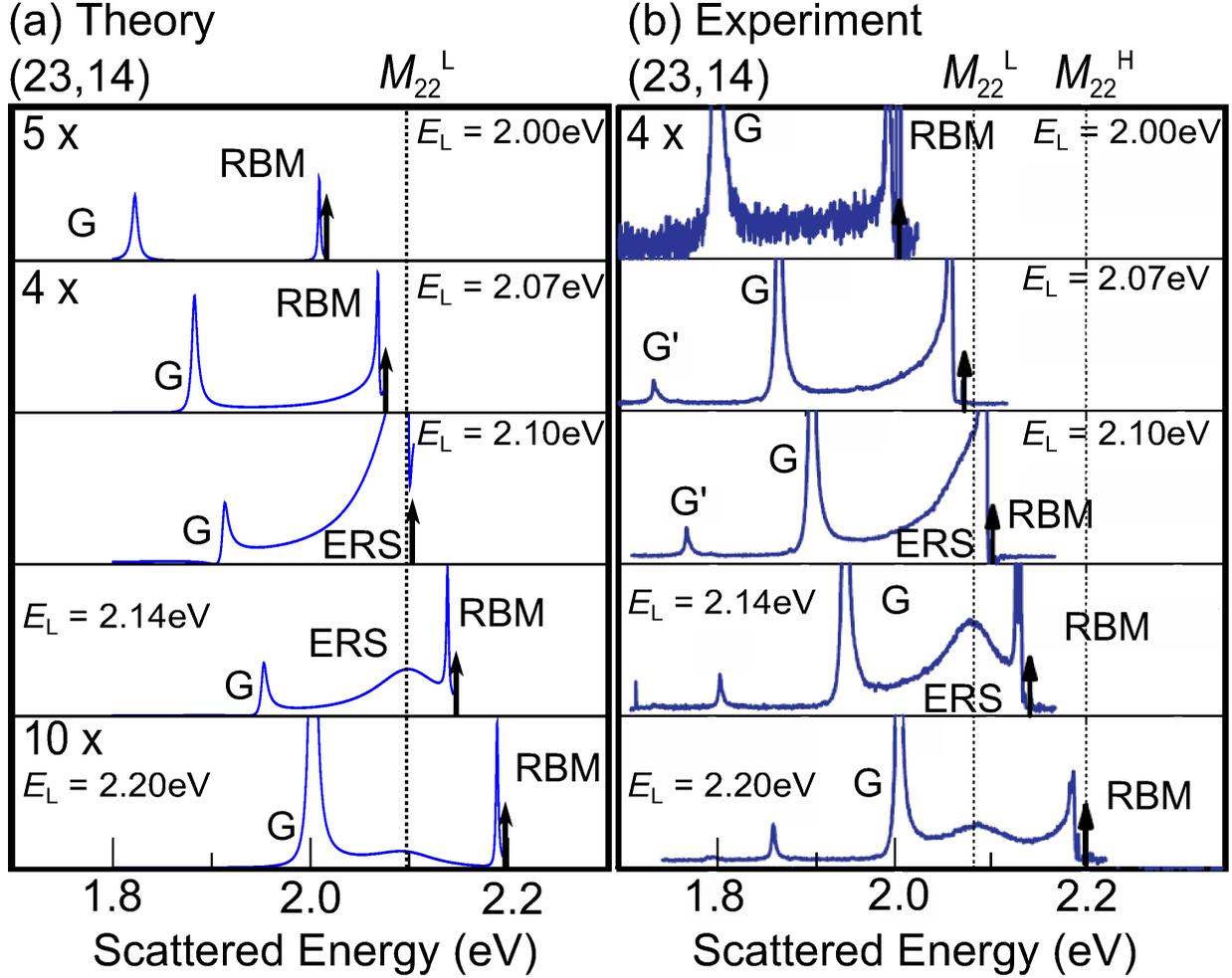}
\caption{\label{fig:Calculated-and-Experimental} (Color online) (a)
  Calculated result (this work) and (b) experimental results (adapted
  from Ref. \cite{PhysRevLett.107.157401}) of Raman intensity versus
  scattered photon energy $\left(\hbar \omega_{\mathrm{s}}\right)$ for
  a (23,14) tube where we have the calculated $M_{22}^{L}=2.10$ eV and
  the experimental $M_{22}^{L}=2.08$ eV.  The laser excitation
  energies $E_{\rm L}$ are taken as 2.00, 2.07, 2.10, 2.14, and 2.20
  eV.}
\end{figure}

Next, to explain the Fano resonance in m-SWNTs, we calculate the Raman
intensity by taking into account each contribution from the RBM, ERS,
and G band, and considering all possible initial ($i$) and final ($f$)
states:
\begin{equation}
  I =\sum_{i}\left|\sum_{f}\left(A_{\mathrm{RBM}}+
      A_{\mathrm{ERS}}+A_{\mathrm{G}}\right)\right|^{2}, \label{eq:Intens}
\end{equation}
in which we have $A_{\mathrm{RBM}}$ and $A_{\mathrm{G}}$ for the
phonon spectral amplitudes and $A_{\mathrm{ERS}}$ for the electronic
scattering amplitude. We do not consider the G' band because its
position ($\sim$ 2700 $\rm{cm}^{-1}$) is quite far from the ERS  and
might not interfere with the ERS as indicated in Farhat's experiment
(Fig.~\ref{fig:Calculated-and-Experimental}(b)).  The amplitude of
each phonon spectrum can be calculated by:
\begin{align}
  A_{\nu}(\omega_{\mathrm{s}}) = & \frac {1}{\pi} \sum_{n,n'} \bigg[ 
  \frac{{\cal M}_{\rm ex-op}^{n,i}} {\left[\Delta
      E_{ni}-i\gamma\right]} \frac{{\cal M}_{\rm ex-ph}^{n',n}}
  {\left[\Delta
      E_{n'i}-\hbar\omega_{\nu}-i(\gamma+\Gamma_{\nu})\right]}\nonumber\\
  &\times \frac{{\cal M}_{\rm ex-op}^{f,n'}}
  {\left[E_{\mathrm{L}}-\hbar\omega_{\nu}-\hbar \omega_\mathrm{s}
      -i\Gamma_{\nu}\right]} \bigg], \label{eq:amplitude}
\end{align}
where $\nu$ = RBM or G mode, $\Delta E_{mi}=E_{\rm L}-E_{m}-E_{i}$,
and $\hbar \omega_{\mathrm{s}}$ is the scattered photon energy.  We
use a broadening factor $\gamma$ = 60 meV for the life time of the
photo-excited carriers~\cite{Sato201094}.  We also utilize the phonon
spectral width for the RBM as a constant $\Gamma_{\mathrm{RBM}}=10\
\mathrm{cm^{-1}}$, and for the G band, which consist of in-plane
transverse optic (iTO) $\Gamma_{\mathrm{iTO}} = 20\ \mathrm{cm^{-1}}$
and longitudinal optic (LO) $\Gamma_{\mathrm{LO}} = 31\
\mathrm{cm^{-1}} $ \cite{Lazzeri-no-plasmon}.  The exciton-photon
(${\cal M}_{\rm ex-op}^{b,a}$) and exciton-phonon (${\cal M}_{\rm
  ex-ph}^{b,a}$) matrix elements for a transition between states $a
\rightarrow b$ are taken from Jiang's work~\cite{PhysRevB.75.035405}.
We approximate the virtual states $i=f$ and $n=n'$.  On the other
hand, the amplitude of the second-order ERS process is given by:
\begin{align}
  A_{\mathrm{ERS}}(\omega_{\mathrm{s}}) =&
  \frac{1}{\pi}\sum_{n,n',n'',\sigma} \bigg[ \frac{{\cal M}_{\rm
      ex-op}^{n,i}}{\left[\Delta E_{ni}-i\gamma\right]} \nonumber\\
  &\times \frac{{\cal M}_{\rm ex-ex}^{n',n} (q)} {\left [\Delta
      E_{n'i}-\hbar\omega_{1} - i(\gamma+\Gamma_{\mathrm{x}})
    \right]} \nonumber\\
  &\times \frac{{\cal M}_{\rm ex-ex}^{n'',n'}(-q)} {\left[\Delta
      E_{n''i}-\hbar\omega_{1}-\hbar\omega_{2}-i(\gamma+
      2\Gamma_{\mathrm{x}})\right]}\nonumber\\
  &\times \frac{{\cal M}_{\rm ex-op}^{f,n'}}
  {\left[E_{\mathrm{L}}-\hbar\omega_{1}-\hbar\omega_{2} -
      \hbar\omega_{\mathrm{s}}-2i\Gamma_{\mathrm{x}}\right]} \bigg],
\end{align}
where we also consider the same virtual state approximation as in
Eq. \eqref{eq:amplitude}.  Here, $\omega_{1}$ and $\omega_{2}$ are the
energies of the linear band excitons emitted from the exciton-exciton
interaction in the second-order ERS process.  The summation over
$\sigma$ denotes all different processes in the ERS mechanism, i.e. AV
and EV scattering processes.  The electron-electron interaction life
time is set at a constant value $\Gamma_{\mathrm{x}}=$ 25 meV.

\begin{figure}
\includegraphics[width=\columnwidth]{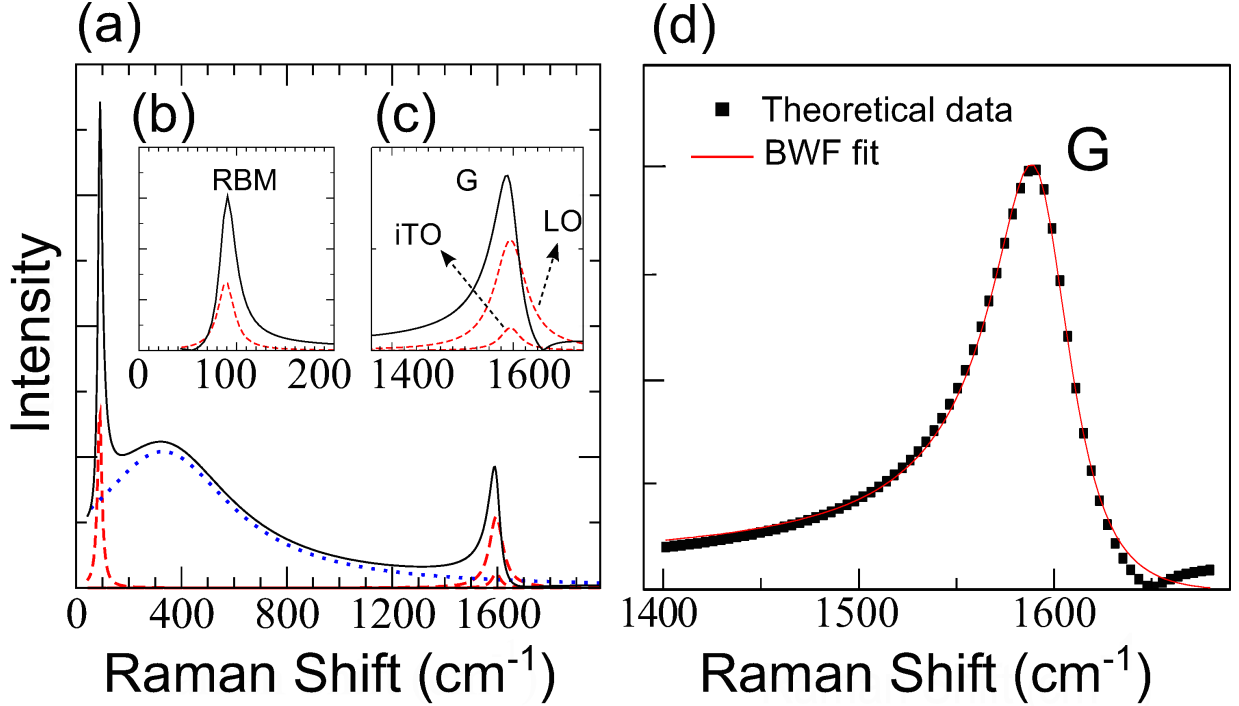}
\caption{ (Color online) Calculated Raman spectra for a (23,14) SWNT with
  $E_{\rm L}= 2.14$ eV.  The total intensity shown in panel (a) is
  represented by the solid line.  The dashed lines show contributions
  from the RBM and G modes, while the dotted blue line is the
  contribution from the ERS.  Each line shape for the RBM, the G modes,
  and the ERS are Lorentzian. (b) The RBM and (c) the G band spectra
  after subtracting the ERS spectrum. The dashed lines are identical
  with those in (a). (d) Fitting result of the G band spectrum to the
  BWF line shape of Eq. \eqref{bwf}.  Filled squares are calculated
  results and the solid line shows the BWF fitting.\label{fig4}}
\end{figure}

In Fig. \ref{fig:Calculated-and-Experimental}(a) we show the
calculated result of the $E_{\mathrm{L}}$ dependence of the Raman
intensity as a function of scattered photon energy $\left(\hbar
  \omega_{\mathrm{s}}\right)$.  In the present work, we only calculate
the $E_{\rm{L}}$ dependence of the Raman intensity near $M_{22}^{\rm
  L}$. Despite the two contributions from the LO and iTO phonon
vibrations for the G mode, the splitting of the $\rm{G}^+$ and
$\rm{G}^-$ modes do not appear visibly in the spectra due to the large
diameter of the (23,14) tube ($d_t = 2.5$ nm) which makes it has a
small curvature effect.  Even though we can not reproduce the relative
intensity scale exactly from the experimental data, our calculated
result can explain the behavior of the observed ERS as shown in
Fig. \ref{fig:Calculated-and-Experimental}(b).  The ERS feature has a
very broad spectral width ($\mathrm{FWHM}^{\mathrm{ERS}}\approx50$
meV) with a peak intensity almost comparable to that of the RBM.
Unlike the other phonon modes, whose peak positions are shifted by
changing $E_{\rm{L}}$, the ERS peak remains at the frequency of the
$M_{ii}$ transition.  At $E_{\mathrm{L}} = 2.07$ eV, the ERS spectrum
starts to appear and modifies the RBM and the G band line shapes.  At
that point, although $E_{\rm{L}}$ is $30$ meV below $M_{ii}$, the
energy-momentum conservation during the exciton-exciton scattering may be
violated by the Heisenberg uncertainty principle ($\Delta t\approx 10$
fs corresponding to $\Delta E\approx 100$ meV).

\begin{figure}
\includegraphics[width=\columnwidth]{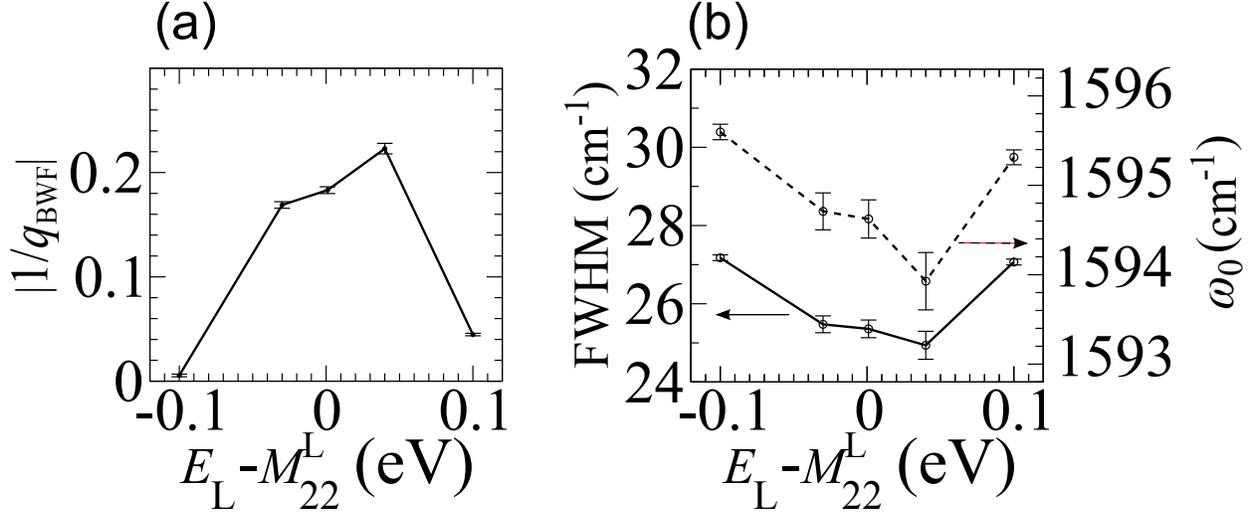}
\caption{(a) Asymmetric factor
($1/q_{\mathrm{BWF}}$), and (b) spectral
width and peak position of the G band as a function of
resonance condition for the (23,14) tube.  The solid and dashed arrows
are given as a guide for the corresponding axes.
\label{fig5}.}
\end{figure}

Each Raman intensity calculated from Eq.~\eqref{eq:Intens} actually
gives a Lorentzian shape for all phonon modes and also for the ERS as
presented in Fig. \ref{fig4}(a).  However, the broad feature of the
ERS overlaps with the phonon modes and thus the interference between them
gives rise to the asymmetric line shape, peak shifting, and
the enhancement of both the RBM and the G bands, which can be seen in
Figs. \ref{fig4}(b) and (c).  We find that the asymmetric line shape
of the G band after subtracting the ERS contribution clearly shows
the BWF line shape (Fig. \ref{fig4}(d)), fitted by
\begin{equation}
  I\left(\omega\right) = I_{0}\frac{\left[1 + \left(\omega -
        \omega_{0}\right) / q_{\mathrm{BWF}}\Gamma\right]^{2}}{1 +
    \left[ \left(\omega-\omega_{0}\right)/\Gamma\right]^{2}},\label{bwf}
\end{equation}
where $q_{\rm BWF}$, $\Gamma$, and $\omega_0$ are parameters to be
determined.  From this fitting, we can find and analyze the
$E_{\rm{L}}$ dependence of the asymmetric factor $1/q_{\mathrm{BWF}}$,
the spectral width $\Gamma$, and the peak position $\omega_{0}$ (see
Fig.~\ref{fig5}).  According to Fano~\cite{PhysRev.124.1866},
$1/q_{\mathrm{BWF}}$ is proportional to the coupling constant between
the continuum spectrum and the discrete spectrum.  In our case,
$|1/q_{\mathrm{BWF}}|$ (FWHM or $\omega_0$) as a function of resonance
condition $E_{\mathrm{L}}-M_{22}^L$ has a ``$\Lambda$'' (``V'') shape,
with the maximum (minimum) peak $\sim$ 40 meV above the resonance as
depicted in Fig. \ref{fig5}(a)$\big($Fig.~\ref{fig5}(b)$\big)$.
$|1/q_{\mathrm{BWF}}|$ reaches a maximum value because the intensity
and the peak position of the ERS allows it to have a very strong
overlap with the G band at that point. This coupling also induces the
narrowing and the shifting of the G band peak closer to the ERS peak
position.

From our theoretical point of view, we suggest some conditions how the
ERS and asymmetric phonon modes in m-SWNTs can be observed
experimentally.  Since the Coulomb interaction is inversely
proportional to the SWNT diameter $d_t$ and also sense a curvature-induced
band gap ($\sim \cos \theta/d_t^2$ meV) appears for  small chiral angles
$\theta$ for $ d_t <$ 1 nm~\cite{Ouyang27042001}, the diameter range
of m-SWNTs which allows us to observe the ERS and Fano resonance
should be around $1-2$ nm.  Moreover, the finite length of m-SWNTs
leads to discrete $k$ points and the electron-electron interaction
energy is around $60$ meV; thus the nanotube length should be larger
than $4$ $\mu$m for $1$ meV energy resolution. The energy of the
second-order exciton-exciton interaction ($\hbar \omega_1$) is only
$\sim 10$ meV lower than the first-order process because the linear
band slope is steeper than that for the parabolic band. Therefore, in
order to identify the dominant contribution of the second-order
process, the low temperature ($10-100$ K) gate voltage experiment must
be performed.
 
In summary, we have formulated a theoretical picture of the ERS by
considering the exciton-exciton interaction.  We showed that the
non-zero momentum transfer process ${\bf q}\neq 0$ gives a dominant
contribution to the ERS spectra.  This ERS spectrum is strongly
coupled with the G band and the RBM and the interference with the ERS
spectrum modifies the line shapes of the two phonon modes which
results in these phonon modes having the BWF line shapes.  The
asymmetry, narrowing, and shifting of the G band induced by
interference with the ERS are all sensitive to the peak intensity
ratio and the peak distance between the ERS and the G band.  The RBM
mode also is predicted to have a similar asymmetry which opens up the
possibility for future experimental observations and clarifications.

E.H. and A.R.T.N. are supported by a MEXT scholarship. R.S. and K.S.
acknowledge MEXT grant Nos. 20241023 and 23710118,
respectively. M.S.D. acknowledges NSF-DMR grant No. 10-04147.

\bibliographystyle{aip}

\end{document}